\documentclass[altaffilletter,superscriptaddress,amsmath,amssymb,twocolumn]{revtex4-1}
\usepackage{graphicx}
\usepackage{textcomp}
\usepackage{natbib}

\begin{document}

\title{Ultra-broadband quantum cascade laser operating from 1.88 to 3.82 THz}%

\author{Markus R\"osch}%
\email{mroesch@phys.ethz.ch}
\author{Mattias Beck}
\author{Martin J. S\"uess}
\affiliation{ETH Zurich, Institute of Quantum Electronics, Auguste-Piccard-Hof 1, 8093 Zurich, Switzerland}
\author{Dominic Bachmann}
\author{Karl Unterrainer}
\affiliation{TU Wien, Photonics Institute and Center for Micro- and Nanostructures, Gu{\ss}hausstra{\ss}e 27-29, 1040 Vienna, Austria}
\author{J\'er\^ome Faist}
\author{Giacomo Scalari}
\email{scalari@phys.ethz.ch}
\affiliation{ETH Zurich, Institute of Quantum Electronics, Auguste-Piccard-Hof 1, 8093 Zurich, Switzerland}

\begin{abstract}
We report on a heterogeneous active region design for terahertz quantum cascade laser based frequency combs. Dynamic range, spectral bandwidth as well as output power have been significantly improved with respect to previous designs. When operating individually the lasers act as a frequency comb up to a spectral bandwidth of 1.1 THz, while in a dispersed regime a bandwidth up to 1.94 THz at a center frequency of 3 THz can be reached. A self-detected dual-comb setup has been used to verify the frequency comb nature of the lasers. 
\end{abstract}

\maketitle 

\begin{figure}[tb]
\centering
  \includegraphics[scale=1]{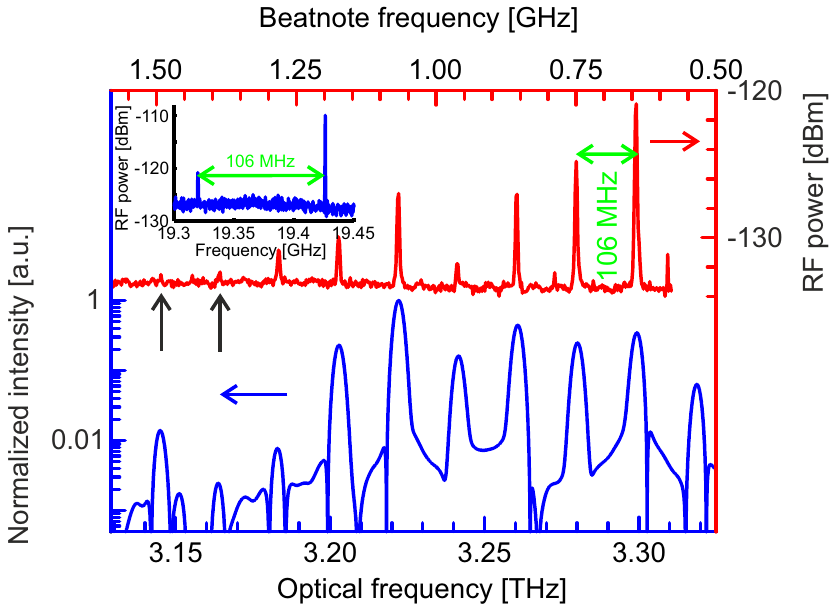}
  \caption{Dual-comb experiment: FTIR (Bruker 80v) measurement of two lasers operating as FCs (blue curve). The two lasers are both 2\,mm x 60\,\textmu m. Laser 1 is driven at 595\,mA and laser 2 at 550\,mA. The temperature is fixed to 26 Kelvin. Only a single set of modes is observed as the offset between the two sets of laser modes is below the resolution limit of the FTIR (2.25\,GHz). The red curve is the corresponding multiheterodyne spectrum recorded with an RF spectrum analyser (Rohde \& Schwarz FSU50; RBW: 2\,kHz, VBW: 20\,kHz, SWT: 250 sec.). The signal was extracted with a bias-tee from the bias line of one of the two lasers. The two black arrows indicate the positions of the two weakest modes of the multiheterodyne spectrum. The inset shows the two intermode beatnotes of the two lasers (RBW: 10\,kHz, VBW: 100\,kHz, SWT: 2 sec.).}
  \label{fig:HDBN}
\end{figure}

The quantum cascade laser (QCL) is an established coherent source for wavelengths between 3\,\textmu m and 300\,\textmu m \cite{faist1994,williams2007,scalari2008review,revin2011_3um}. Especially in the so-called terahertz (THz) frequency range (roughly 1-10\,THz or 30-300\,\textmu m) QCLs are a compact direct source providing high power up to the watt level \cite{williams2007,li2014highpower}. Due to its inherent properties QCLs can be designed to cover a very broad spectral range with a single device \cite{gmachl2002,turcinkova2011,Roesch2014}, making them an attractive source for broadband spectroscopy. \\
Frequency combs (FC) have recently been reported as a tool for spectroscopic applications with QCLs \cite{hugi2012,Burghoff2014,Roesch2014}. So-called dual-comb spectroscopy is particularly interesting at both midinfrared and THz frequencies \cite{Bernhardt2010,baumann2011spectroscopy,zhang2013,yasui2006,finneran2015,villares2014,Roesch2016,yang2016}. To achieve FC operation using a THz QCL the dispersion of the laser has to be engineered to be close to zero for the frequency range of operation \cite{Burghoff2014,Roesch2014}. Two different approaches have been reported to achieve low and flat dispersion in THz QCLs. The first one introduces negative dispersion to compensate for the positive intrinsic dispersion of narrow waveguides with a double-chirped mirror \cite{Burghoff2014}. The alternative approach uses wider waveguides in combination with a broad gain medium which provides a low and flat intrinsic dispersion at the center of the gain curve \cite{Roesch2014,Roesch2016,bachmann2016dispersion}. The spectral range of the FC is in that case directly related to the gain curve \cite{bachmann2016dispersion}. A wider gain curve will allow FC operation as long as the gain curve remains flat \cite{Roesch2014,bachmann2016dispersion}. However, it needs to be noted that a FC operation on the full laser bandwidth will only be possible by artificially introducing additional dispersion i.e. by a Gires-Tournois interferometer \cite{gires1964interferometre,villares2016gti}, or a double-chirped mirror \cite{szipocs1994,Burghoff2014}. \\
In this work, we present an active region design based on a heterogeneous cascade structure that expands both the total laser bandwidth as well as the FC bandwidth. This shows that indeed by gain engineering the dispersion can be further compensated and larger FC bandwidths can be achieved. In addition, also the dynamic range as well as the output power are increased. An experiment similar to the one reported in Ref. \cite{Roesch2016} has been used to verify the FC operation in a dual-comb configuration. \\
The laser active region we present here is based on the design reported in Ref. \cite{Roesch2014}, which has been further developed. It fully exploits the capability of QCLs to integrate different active region designs within one laser cavity \cite{gmachl2002}. The used building block is a design which by itself is already intrinsically broadband, a four quantum-well design reported originally in Ref. \cite{amanti2009}. This active region was then adapted to have four designs with slightly different central frequencies. The four designs have central frequencies of 2.3, 2.6, 2.9, and 3.4 THz. While the three lower frequency designs are identical with those from Ref. \cite{Roesch2014}, the design at 3.4 THz has been added to increase the bandwidth towards higher frequencies. The number of periods per design has also been rearranged in order to provide a flat gain resulting in a similar threshold for all the active regions and more dynamic range. Additionally, the doping level has been increased to $2.2 \times 10^{16}$ cm$^{-3}$. The exact layer sequences are reported in the appendix.\\
\begin{figure*}[htb]
  \centering
  \includegraphics[scale=1]{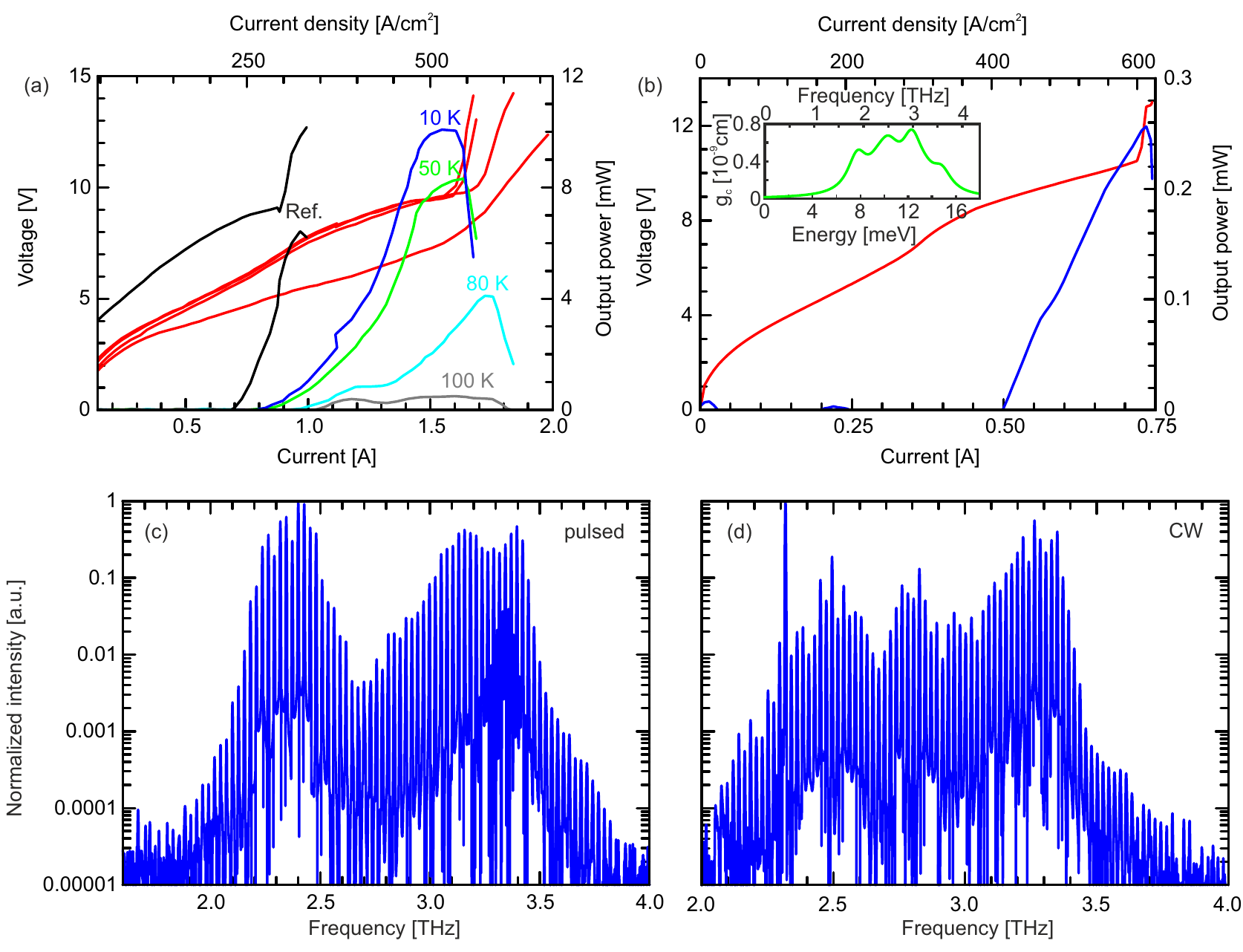}
  \caption{Laser characteristics: (a) Light-current and current-voltage characteristics for a 2\,mm x 150\,\textmu m wet-etched laser in pulsed operation at different temperatures. To emphasize the improvements a reference measurement of a laser with the same dimensions (black curves) based on the design reported in Ref. \cite{Roesch2014} is shown. The measurements were performed in pulsed operation with 10\% duty cycle (5\% for the 100\,Kelvin measurement). The optical power was measured with a calibrated powermeter (TK instruments). (b) Light-current and current-voltage characteristics for a 2\,mm x 60\textmu m dry-etched laser in CW operation at 20 Kelvin. The power was measured with a calibrated powermeter (Ophir:3A-P-THz). The inset shows the calculated gain cross-section $g_c$ of the reported laser design. (c) Optical spectrum of a 1.5\,mm x 150\,\textmu m wet-etched laser in pulsed operation (20\,\% duty cycle) at 20 Kelvin. (d) Optical spectrum of a 1.8\,mm x 60\,\textmu m dry-etched laser in CW operation at 21 Kelvin. The spectra in (c) and (d) were measured using a commercial vacuum FTIR (Bruker v80).}
  \label{fig:liv}
\end{figure*}
Lasers were fabricated into metal-metal waveguides for best performance \cite{williams2003,Roesch2014,bachmann2016setback}. Both wet-etching and dry-etching techniques have been used to define the laser ridges. Wet-etched lasers typically provide the best results in pulsed operation while dry-etched lasers can be fabricated into narrower ridges, and therefore have reduced Joule-heating for the same cavity length which favors continuous wave (CW) operation \cite{bachmann2016setback}. \\
Figure \ref{fig:liv}(a) shows the light-current and current-voltage (LIV) characteristics of a 2\,mm x 150\,\textmu m wet-etched laser in pulsed operation. The laser shows up to 10\,mW of output power at 10 Kelvin. Lasing occurs up to a temperature of 100 Kelvin. For comparison, the LIV of a laser with the same dimensions of the design reported in Ref. \cite{Roesch2014} is shown along in figure \ref{fig:liv}(a). We observe roughly 40\% more power on the reported design. Additionally, the dynamic range of the laser is significantly improved. While for the reference laser of Ref. \cite{Roesch2014} a dynamic range ($J_{max}/J_{th}$) of 1.4 was measured, the reported design shows a dynamic range of 2. We attribute the improvement of the dynamic range to the careful rearrangement of the active region while the improvement in output power is due to a combination of the higher doping level as well as the increased number of cascades in the active region. The higher doping level also causes a slightly higher threshold current.\\
\begin{figure*}[tbh]
  \centering
  \includegraphics[scale=1]{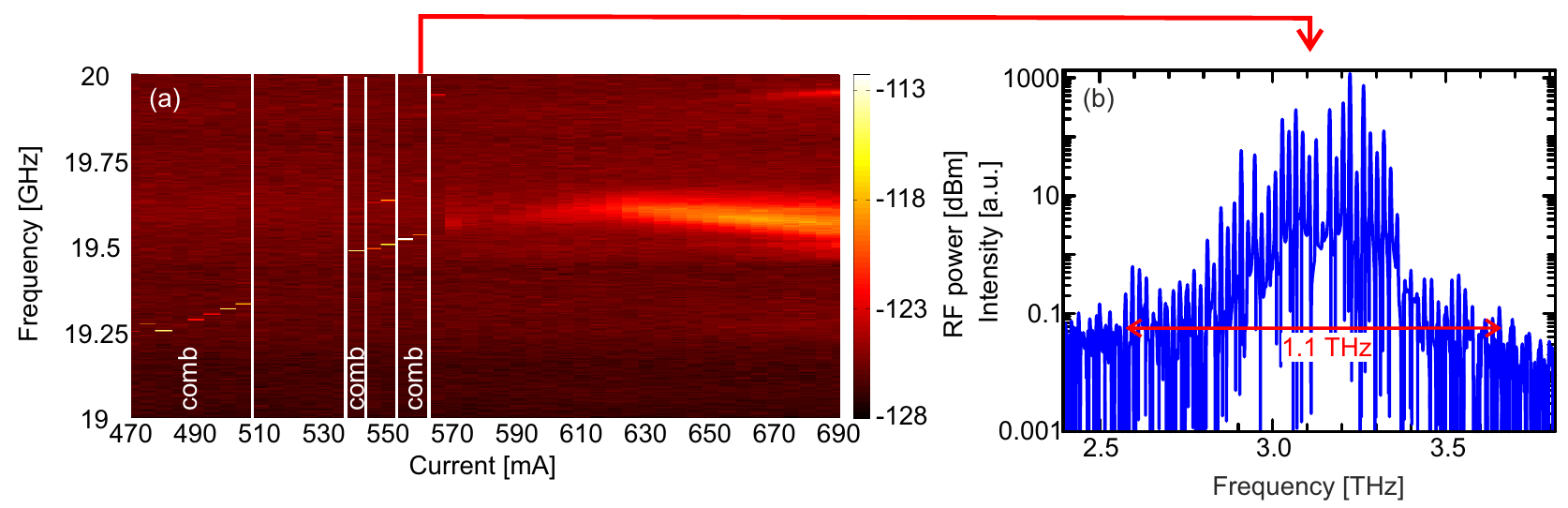}
  \caption{Frequency comb performance: (a) Intermode beatnote of a 2\,mm x 60\,\textmu m dry-etched laser as a function of the driving current. The temperature is fixed to 20 Kelvin. The beatnote signal is extracted from the bias line with a bias-tee and is recorded with an RF spectrum analyser (Rohde \& Schwarz FSU50; resolution bandwidth (RBW): 10\,kHz, video bandwidth (VBW): 100\,kHz, sweep time (SWT): 10 sec.). (b) Optical spectrum for the same laser at 562\,mA at 16 Kelvin. For this current the laser is in a FC regime providing a bandwidth of 1.1\,THz.}
  \label{fig:BNmap}
\end{figure*}
Dry-etched lasers were used to testify the CW performance of the reported active region. Figure \ref{fig:liv}(b) shows the LIV characteristics for a 2\,mm x 60\,\textmu m dry-etched laser in CW operation at 20 Kelvin. An optical power of 0.26\,mW was measured using a calibrated powermeter (Ophir:3A-P-THz). Operation up to a temperature of 30 Kelvin is achieved in CW operation. \\ 
Since the reported laser features an additional active region design with a central frequency at 3.4\,THz the optical spectrum has an upper limit at almost 4\,THz as can be seen in figure
\ref{fig:liv}(c). The spectra were measured using a commercial under-vacuum Fourier-transform infrared spectrometer (FTIR). In total, the lasing spectrum spans over 1.94 THz from 1.88 THz to 3.82 THz covering more than a full octave in frequency. The spectrum is in good agreement with the calculated gain cross-section (inset of figure \ref{fig:liv}(b)) using the same model as in Ref.\cite{Roesch2014}. A similar performance can also be observed in CW for dry-etched lasers. As shown in figure \ref{fig:liv}(d) the lowest frequencies below 2\,THz do not reach lasing threshold in dry-etched lasers. Most likely this is due to increased waveguide losses of the narrow dry-etched waveguide (60\,\textmu m) at low frequencies in combination with the large Joule heating which also prevents the laser from reaching high operation temperatures in CW operation. \\
It has recently been shown that THz QCLs with a broad spectral gain have sufficiently low and flat dispersion to operate as FC for a limited part of the laser's dynamic range \cite{Roesch2014,Roesch2016}. Similar behavior can be expected from the reported laser design as it covers an even larger spectral part. The key indicator of comb operation in QCLs is the radio frequency (RF) beatnote signal at the roundtrip frequency of the laser cavity ($f_{rep}$). A single narrow beatnote at $f_{rep}$ is indicating comb operation while a broad beatnote is characteristic for a lasing regime where the group velocity dispersion (GVD) is large enough to prevent the four-wave mixing to lock the lasing modes \cite{hugi2012,Khurgin2014,Burghoff2014,Roesch2014,Roesch2016}. Throughout the entire dynamic range of the laser the RF beatnote around the roundtrip frequency of the laser is recorded using a bias-tee on the bias line and a radio frequency spectrum analyser (Rohde \& Schwarz FSU50). The corresponding beatnote map for a 2\,mm x 60\,\textmu m dry-etched laser is shown in figure \ref{fig:BNmap}(a). The performance is similar to the one reported in Ref. \cite{Roesch2014}. For lower currents a single narrow beatnote is observed while for higher currents dispersion sets in preventing comb operation, and resulting in a broad beatnote. \\
Figure \ref{fig:BNmap}(b) shows the optical spectrum at the highest bias point with a narrow beatnote (562\,mA). A spectral bandwidth of 1.1\,THz is recorded at a central frequency of 3.1\,THz. To our knowledge this is the broadest THz QCL FC so far reported. \\
A narrow beatnote at the roundtrip frequency of the laser is a necessary but not a sufficient condition to prove FC operation \cite{Burghoff2014,Roesch2014,Roesch2016}. Beatnote spectroscopy or shifted wave interference Fourier transform spectroscopy experiments that prove the comb operation are difficult to perform at THz frequencies due to the lack of fast and sensitive enough detectors \cite{hugi2012,Burghoff2014}. Another approach to prove the comb behavior of a QCL comb is to perform a dual-comb experiment \cite{villares2014,Roesch2016}. Such an experiment will directly reveal the comb characteristics of the two lasers used for the experiment. For THz QCL combs a self-detection dual-comb configuration allows a straight-forward implementation of that experiment without the need of an external detector \cite{Roesch2016}. \\
Two lasers mounted on the same chip parallel to each other have been used to perform this experiment in analogy to Ref. \cite{Roesch2016}. The two lasers have the dimensions of 2\,mm x 60\,\textmu m  and were driven in CW operation at a temperature of 26 Kelvin. A dual-comb regime was identified in a first step by the presence of two narrow beatnotes at the roundtrip frequencies of the two lasers (see inset of figure \ref{fig:HDBN}). For the same driving conditions the optical spectrum of the two lasers was recorded with an FTIR. The corresponding optical spectrum in figure \ref{fig:HDBN} (blue curve) shows a single set of equally spaced modes. This hints that the offset between the two sets of comb modes is lower than the resolution of the FTIR (2.25\,GHz). Indeed the multiheterodyne spectrum is located between 0.5 and 1.5\,GHz (red curve in figure \ref{fig:HDBN}). Nine modes can be identified with a mode spacing of 106\,MHz in good agreement with the spacing of the intermode beatnotes as well as the recorded optical spectrum. \\
The multiheterodyne spectrum corresponds to an optical bandwidth of 175\,GHz. This is significantly less than the reported bandwidth for a single device in figure \ref{fig:BNmap}(b). This can be attributed to the large Joule heating when two laser are driven in CW operation at the same time on the same cryostat. For the same reason it was not possible to cool down the two lasers to temperatures below 25 Kelvin while for a single laser operation temperatures down to 15 Kelvin can be achieved with the same Helium-flow cryostat. Therefore the lasers had to be operated close to their maximal operation temperature which significantly decreases the spectral bandwidth. However the measurements in figure \ref{fig:HDBN} clearly show that the reported laser is indeed working as a FC supporting the FC identification through the beatnote measurements in figure \ref{fig:BNmap}(a). \\
In summary, we presented a heterogeneous cascade active region design for THz QCLs based on four active region designs. The lasers show a spectral bandwidth of almost 2\,THz at a central frequency of 3\,THz. A peak output power of 10\,mW in pulsed operation was achieved with a dynamic range of $J_{max}/J_{th}=2$. 
FC operation up to a spectral bandwidth of 1.1\,THz was measured. To confirm FC operation a self-detected dual-comb experiment was performed. We could confirm the FC operation on a bandwidth of 175\,GHz. The bandwidth in this experiment was temperature limited, and therefore only a limited part of the effective FC spectrum could be probed. To conclude, we could show that with careful designing and balancing of the involved active region designs the spectra of THz QCLs can be further pushed to bandwidths exceeding an octave in frequency. Ideally such a further improvement will lead to lasers where an octave bandwidth within a 20\,dB power range can be reached. Such power constraints will be necessary for a potential f-to-2f stabilization in QCL FCs \cite{telle1999,jones2000f2f}. \\

\subsection*{Acknowledgement}
The presented work was funded by the EU research project TERACOMB (Call identifier FP7-ICT-2011-C, Project No.296500), and the Swiss National Science Foundation (SNF) grant 200020 165639. The funding is gratefully acknowledged. The authors acknowledge the joint cleanroom facility FIRST at ETH Zurich and the Center for Micro- and Nanostructures (ZMNS) at TU Wien for sample processing.\\

\subsection*{Appendix}
\textbf{Quantum cascade layer sequence and details}.\\
Sample EV2172 has been grown by molecular beam epitaxy (MBE) on a semi-insulating GaAs substrate in the GaAs/AlGaAs material systems. The layer sequence for the 3.5 THz active region (64 repetitions) is, starting from the injection barrier: ${\bf 5.5}/10.7/{\bf 1.4}/10.1/{\bf 3.8}/9.2/{\bf 4.2}/18.0$. 
The figures in bold face represent the Al$_{0.15}$Ga$_{0.85}$As barrier  and the 18.0 nm GaAs quantum well is homogeneously Si doped $2.2 \times 10^{16}$ cm$^{-3}$. The layer sequence for the 2.9 THz active region (34 repetitions) is, starting from the injection barrier: ${\bf 5.5}/11.0/{\bf 1.8}/11.5/{\bf 3.8}/9.4/{\bf 4.2}/18.4$. The 18.4 nm GaAs quantum well is homogeneously Si doped $2.2 \times 10^{16}$ cm$^{-3}$. The layer sequence for the 2.6 THz active region (39 repetitions) is, starting from the injection barrier: ${\bf 5.5}/11.3/{\bf 1.8}/11.3/{\bf 3.8}/9.4/{\bf 4.2}/18.4$. The 18.4 nm GaAs quantum well is homogeneously Si doped $2.2 \times 10^{16}$ cm$^{-3}$. 
The layer sequence for the 2.3 THz active region (73 repetitions) is, starting from the injection barrier: ${\bf 5.5}/12.0/{\bf 1.8}/10.5/{\bf 3.8}/9.4/{\bf 4.2}/18.4$. The 18.4 nm GaAs quantum well is homogeneously Si doped $2.2 \times 10^{16}$ cm$^{-3}$. The three lower frequency layer sequences are identical with those reported in references \cite{turcinkova2011,Roesch2014}. Modifications have been done on the number of repetitions and the Si doping level.

\end{document}